# High luminosity interaction region design for collisions inside high field detector solenoid[1]

**Catia Milardi[2], Miro Andrea Preger, Pantaleo Raimondi and Francesco Sgamma**

*Istituto Nazionale di Fisica Nucleare, Laboratori Nazionali di Frascati,*
*Via Enrico Fermi 40, 00044 Frascati, Roma, Italy.*

*E-mail*: catia.milardi@lnf.infn.it

ABSTRACT: An innovatory interaction region has been recently conceived and realized on the Frascati DAΦNE lepton collider. The concept of tight focusing and small crossing angle adopted to achieve high luminosity in multibunch collisions has evolved towards enhanced beam focusing at the interaction point with large horizontal crossing angle, thanks to a new compensation mechanism for the beam-beam resonances. The novel configuration has been tested with a small detector without solenoidal field yielding a remarkable improvement in terms of peak as well as integrated luminosity. The high luminosity interaction region has now been modified to host a large detector with a strong solenoidal field which significantly perturbs the beam optics introducing new design challenges in terms of interaction region optics design, beam transverse coupling control and beam stay clear requirements. Interaction region design criteria as well as the luminosity results relevant to the structure test are presented and discussed.

# Contents

## 1. Introduction

DAΦNE [1] is the Frascati electron/positron collider working at the c.m. energy of the Φ meson resonance (1020 MeV). It came in operation in 2001 and till summer 2007 provided luminosity, in sequence, to three different experiments which logged a total integrated luminosity of ~ 4.4 fb$^{-1}$: KLOE [2], a multipurpose experiment for the study of K meson decays, as well as hadronic physics and low energy QCD, FINUDA [3], for hypernuclear spectroscopy and DEAR [4] for kaonic atoms.

In 2007 the collider has been upgraded implementing a new collision scheme based on large Piwinski angle and *Crab-Waist* technique for compensation [5] of the beam-beam interaction. The large Piwinski angle, obtained by increasing the horizontal crossing angle and reducing the transverse horizontal beam size at the Interaction Point (IP), provides several advantages [6]: it reduces the beam-beam tune shift in both planes, shrinks the longitudinal size of the overlap areacbetween the colliding bunches, thus allowing for a lower $\beta^*_y$ at the IP, and cancels almost all the parasitic crossings: in fact it becomes possible to completely separate the vacuum chambers of the two beams just after the first low-beta quadrupole in the interaction region (IR). A couple of *Crab-Waist* (CW) sextupoles, installed in symmetric position with a proper phase advance with respect to the IP, suppresses the betatron and sinchrobetatron resonances coming from the vertical motion modulation due to the horizontal oscillation.

The new configuration has been used to provide beam-beam events to the SIDDHARTA [7] experiment a compact device, heir of DEAR, without solenoidal field, the simplest environment for testing the performances of the new approach to collisions. During the test the measured luminosity has been increased by a factor 3 with a peak value of $4.53 \times 10^{32}$ cm$^{-2}$s$^{-1}$ [8,9] letting in collision currents slightly lower than those corresponding to the old records. The highest daily integrated luminosity measured in a moderate injection regime, compatible with the detector requirements, has been $L_{\int day}$ ~15 pb$^{-1}$. An almost continuous injection regime provided $L_{\int 1\ hour}$ ~1.0 pb$^{-1}$ hourly integrated luminosity which opened significant perspectives for large experiment data taking. Scaling this best integrated luminosity measured over two hours, it is reasonable to expect more than 20 pb$^{-1}$ per day, and assuming 80% collider uptime as during the past runs, ~ 0.5 fb$^{-1}$ per month [10].

The results of the high luminosity test renovated the interest in planning and undertaking experimental activities on DAFNE, paving the way for a new run with an upgraded KLOE detector, KLOE-2 [11].

## 2. The high luminosity interaction region

Integrating the high luminosity collision scheme with a large experimental detector introduces new challenges in terms of IR layout and optics, beam acceptance as well as coupling correction [12].

### 2.1 Impact of the experimental detector on the interaction region design

The KLOE-2 detector is equipped with a superconducting solenoid providing an intense 2.3 Tm integrated magnetic field, which, due to the low energy of the machine (Bρ=1.7 Tm), must be taken into account as integral part of the magnetic lattice. It affects the transverse beam dynamics perturbing the beam trajectory, adding focusing effects and transverse beam coupling. Concerning coupling it is worth reminding that the transverse oscillation plane of the beam is rotated by an angle ~39°, in module, in half solenoid. Moreover the new detector setup introduces several modifications posing strict requirements on the IR design in terms of stay-clear aperture. Additional detector layers, including both tracking and calorimeter devices, have been inserted in the inner part of the apparatus, close to the interaction region. A very light tracker, consisting of a cylindrical gas electron multiplier (GEM) detector, has been installed in the space between the drift chamber and the spherical beam pipe. Crystal calorimeters, in front the collider low-β quadrupoles, increase the acceptance for photons emitted under a very low angle, a key issue for some rare decay studies. KLOE-2 will also extend its investigation capabilities to the study of γ-γ reactions, tagging the scattered electron and positron, typical of those events by means of dedicated detectors inserted, one, inside the experimental apparatus and the other after the first dipole in the long arc in each ring.

### 2.2 Interaction region layout and vacuum chamber

The main part of the new DAΦNE IR is the low-β section based on permanent magnet quadrupole doublets. The decision to use permanent magnet elements instead of superconducting tunable ones has been imposed by the need to provide the largest possible free solid angle for the detector.

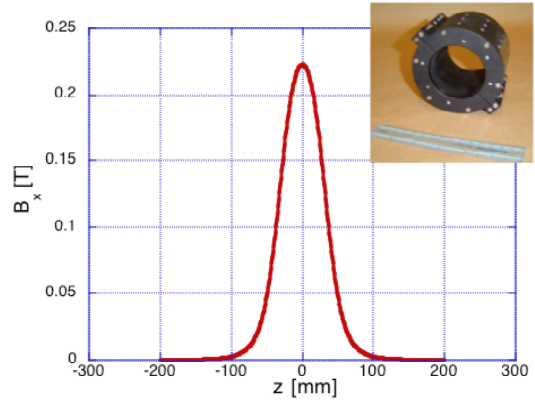

Figure 1. magnetic field of half permanent magnet dipole (top right).

The quadrupoles are made of SmCo alloy and provide gradients of 29.2 T/m for the first one from the IP and 12.6 T/m for the second one. The first (PMQD) is horizontally defocusing and is shared by the two beams; its central azimuthal position has been set at 0.415 m from the IP, as a compromise between the conflicting requirements of tight focusing and large solid angle aperture coming from the collider and the experiment respectively. The second quadrupole (PMQF), horizontally focusing, is installed just after the point where the beam pipes of the two rings are separated and is therefore on axis. Being PMQD much stronger than in the original low- β setup and having doubled the horizontal half crossing angle, now ~25 mrad, a very efficient beam separation, ~ 40 $\sigma_x$, is achieved in the ~1.6 m long section of the IR common to the two rings, making the impact of the two secondary beam encounters occurring on each side of the IP completely negligible. It is worth reminding that usually at DAFNE 100 ÷ 111 contiguous buckets are filled out of the 120 available. Two bunches are separated in time by 2.7 nsec and, as a consequence, only one parassitic crossing is possible in each half of the IR at ~ 41 cm from the IP.

As a drawback the vertical displacement of the beam in the IR, strongly affected by the detector solenoidal field, becomes an order of magnitude larger than in the past.

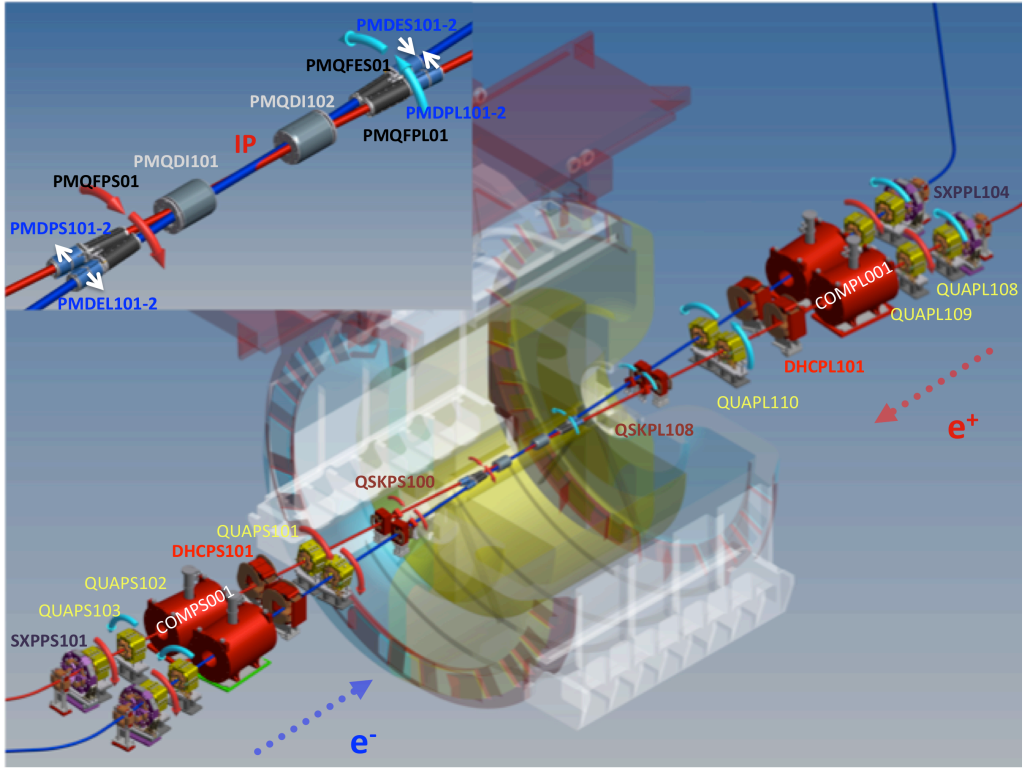

Figure 2. Schematic drawing of the DAFNE interaction region compatible with the KLOE-2 detector and detail of the low-β section (top left).

To keep the beam vertical trajectory within reasonable values a permanent magnet dipole, PMD, has been added just after PMQF, inside the detector magnetic field, in each one of the four IR branches. Each PMD is built by using a SmCo alloy, consists of two parts having 75.0 mm magnetic length each, and provides an integrated field strength BL = 0.0168 Tm, see Fig.1, corresponding to a vertical deflection angle of ~10.0 mrad.

The PMDs are based on a modular design in view of a possible detector run at a lower solenoidal field. They provide a horizontal magnetic field directed inward in the $e^+$ ring and outward in the $e^-$ one, as shown in Fig. 2.

The IR magnetic layout, sketched in Fig. 2, has been designed in order to maximize the beam stay clear treading the beam trajectory as much as possible through the center of the magnetic elements according to a self-consistent procedure. The crossing angle has been tuned in order to have the same horizontal displacement as in the SIDDHARTA configuration at the corrector dipole, DHCPS01, used to match the IR to the ring layout in the arcs, under the constraint of placing the PMDs as close as possible to the PFQMs; its value ($|\theta_c|$ = 25.7 mrad) has been computed switching off PMQFPS01 and the electromagnetic quadrupole QUAPS101 before DHCPS01. The x, x', y, y' coordinates of PMQFPS01 have been calculated imposing that the x, y beam trajectory remains constant at DHCPS01 with or without PMQFPS01 itself. Eventually the DHCPS101 dipole has been used to steer the beam trajectory through the center of the compensator solenoid COMPS001 [13].

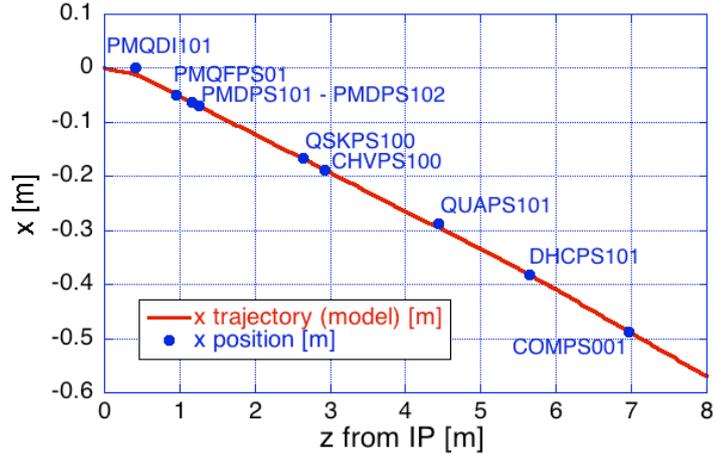

Figure 3. Horizontal trajectory in the IR (solid line) and position of the IR magnetic element centers (dots).

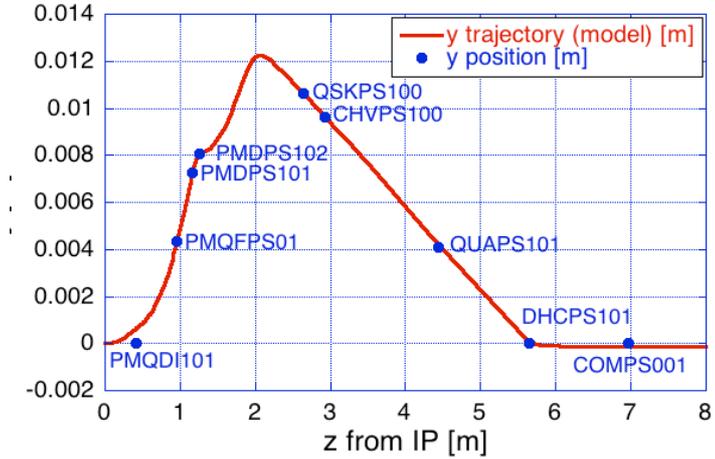

Figure 4. Vertical trajectory in the IR (solid line) and position of the IR magnetic element centers (dots).

The beam trajectory and positions of the magnetic element center positions are shown in Fig. 3 and in Fig. 4 with reference to the IR branch of the positron ring pointing to the short arc, the corresponding branch for the electron ring being symmetric.

The evident advantage of this approach consists in keeping the maximum excursion of the vertical beam trajectory within ~12 mm providing, at the same time, the maximum aperture for the beam.

The horizontal and vertical beam stay-clear requirements have been defined as:

$$X_{SC} = x_{trj} \pm 10\sigma_x$$
$$Y_{SC} = y_{trj} \pm 10\sigma_y$$

where $\sigma_x$ and $\sigma_y$ are the horizontal and vertical rms beam sizes respectively. Their values, computed assuming a conservative large value for the collider emittance $\varepsilon = 0.4 \cdot 10^{-6}$ m for the horizontal plane and full coupling for the vertical one, are presented in Fig. 5 and Fig. 6. Relaying on this analysis the radius of the vacuum pipe, in the section between the IP and the DHCs, has been reduced. It is now 2.75 cm, while it was 4.4 cm during the SIDDHARTA run. The resulting vacuum pipe geometry is largely simplified; in fact it consists of three straight sections, with few junctions and bellows. A narrower vacuum chamber contributes to lower the ring impedance budget, to minimize the strength of trapped high order modes and to shift their frequencies away from the beam spectral lines [14]. The detector efficiency also profits from the larger free space around the IP where a precision vertex tracker can be placed.

The IR pipe is aluminum (AL6082) made with the exception of the sphere surrounding the IP, which is built in ALBEMET. Such a structure could trap HOMs and for this reason it is shielded from to the beam by means of a Be cylinder. To minimise K meson regeneration the shield thickness has been almost halved (35 μm instead of the 65 μm of the last KLOE run).

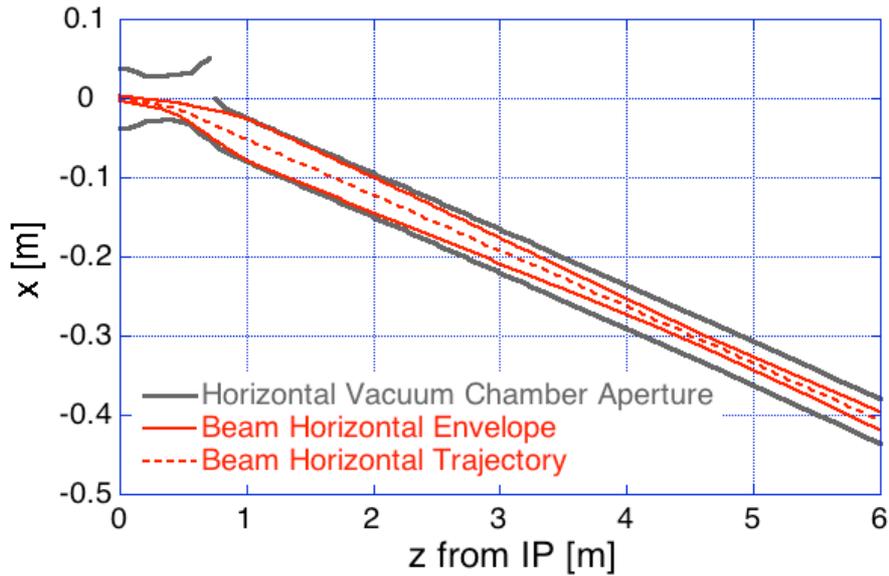

Figure 5. Horizontal beam stay-clear aperture in the KLOE-2 IR.

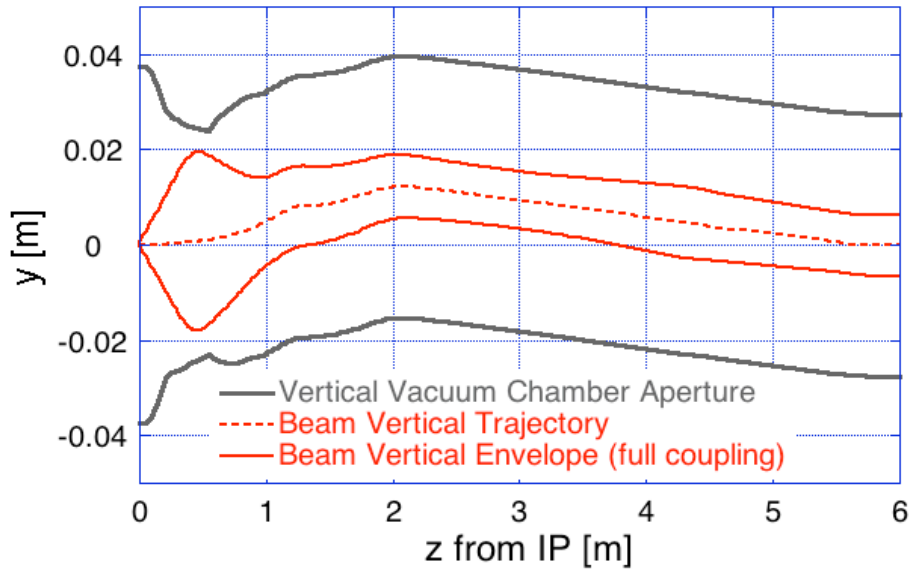

Figure 6. Vertical beam stay-clear aperture in the KLOE-2 IR.

## 2.3 Interaction region optics

The design of the IR optics is constrained by several criteria. It must provide the prescribed low-β parameters at the IP (see Tab. 1), matching at the same time the ring original layout in the arcs. The phase advance between the CW sextupoles and the IP must be π for the horizontal-like mode and π/2 for the vertical one. The values of the β-functions at the CW sextupoles have been designed according to the maximum strength of the existing devices. The transverse coupling introduced by the detector solenoid must be carefully compensated by means of the compensator solenoids and rotation of the IR quadrupoles.

*Table 1: Interaction region parameters*

| | |
|---|---|
| $\beta_x$ [m] | 0.265 |
| $\beta_y$ [m] | 0.0085 |
| $|\theta_c|$ (half value) [rad] | 0.0257 |
| $\alpha_x$ | 0.0 |
| $\alpha_y$ | 0.0 |
| $\eta_x$ [m] | 0.0 |
| $\eta'_x$ | 0.0 |

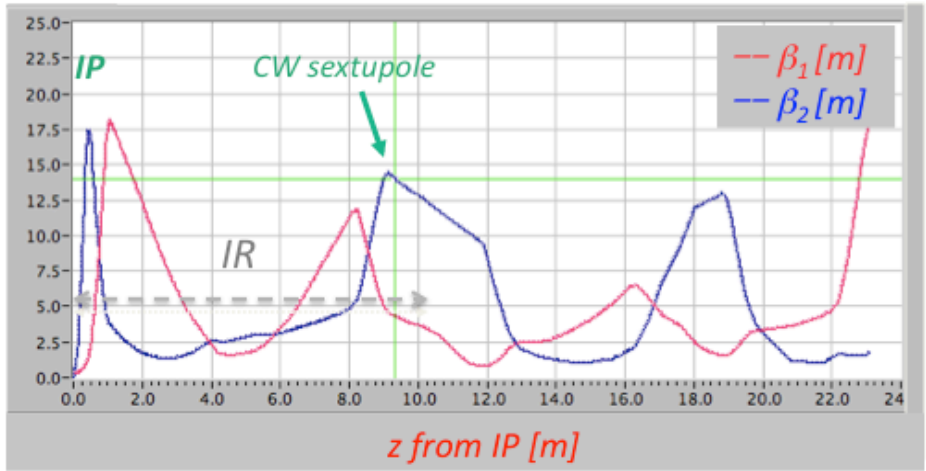

Figure 7. IR optical functions.

## 2.4 Transverse betatron coupling correction

The transverse betatron coupling has been corrected revising the Rotating Frame Method [15] originally used at DAΦNE. The field integral introduced by the solenoidal detector is almost cancelled by means of two anti-solenoids, installed symmetrically with respect to the IP in each ring, which provides compensation also for off-energy particles. The rotation of the beam transverse plane is compensated by co-rotating the quadrupoles PMQFPS01, QUAPS101, QUAPS102, QUAPS103 around their longitudinal axis. The first low beta quadrupole has been kept in the upright position, since the nominal rotation suitable for coupling correction significantly increases the displacement of the beam vertical trajectory (see Fig. 8). The anti-solenoidal field has been set to a value slightly lower than the optimal one in order to minimize the rotation angles and to make the tilts of to the last two electromagnetic quadrupoles antisymmetric. It is worth remarking that the coupling is carefully compensated before the CW sextupoles making the terms of the coupling matrix vanish at QUAPS103

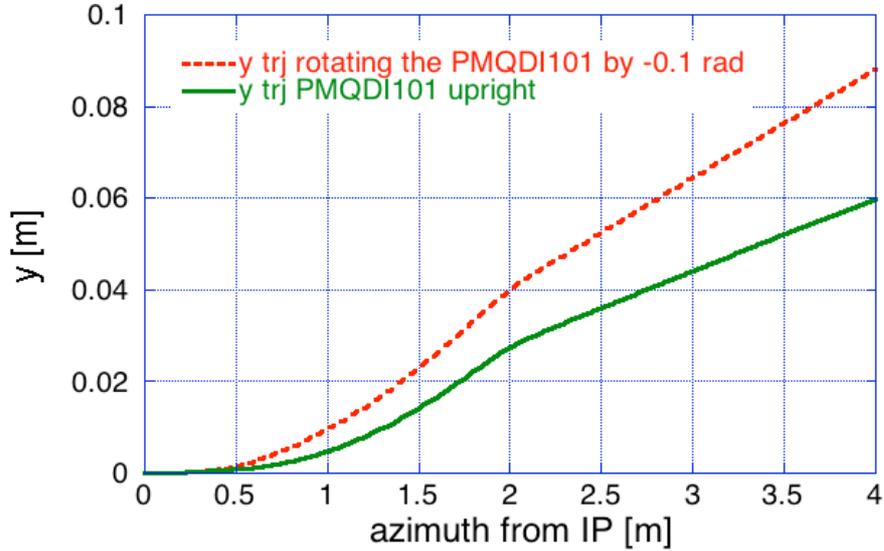

Figure 8. Dependence of the beam vertical trajectory on the nominal rotation of PMQD around its longitudinal axis.

Coupling correction fine-tuning, for each beam, is assured by a skew quadrupole added in each branch of the IR and by the two anti-solenoids independently powered.

The present coupling correction method introduces several advantages with respect to the one adopted for the past KLOE run. The rotation of the permanent magnet quadrupoles installed inside the detector is drastically reduced; in fact it was 8.3° and 12.9° for PMQD and PMQF respectively. Moreover it becomes possible to get rid of the remotely controlled actuators [16] used in the past to finely adjust the low-$\beta$ quadrupole tilts, thus further increasing the solid angle available for the detection of decaying particles. Quadrupole rotations in the four sections of the IR are summarized in Table 2. They are defined so that a positive tilt gives a clockwise rotation in the reference system moving with the positron beam.
The IR parameters are summarized in Tab. 2.

*Table 2. Coupling correction parameters.*

|  | Z from the IP [m] | Quadrupole rotation angles [deg] *Anti-solenoid current [A]* |
|---|---|---|
| PMQDI101 | 0.415 | 0.0 |
| PMQFPS01 | 0.963 | -4.48 |
| QSKPS100 | 2.634 | used for fine tuning |
| QUAPS101 | 4.438 | -13.73 |
| QUAPS102 | 8.219 | 0.906 |
| QUAPS103 | 8.981 | -0.906 |
| *COMPS001* | *6.963* | *72.48 (optimal value 86.7)* |

## 3. Tests with the beam

A full verification of the criteria adopted in designing the IR can be done only letting beams in collision and measuring the convoluted size of the colliding beams at the IP and the specific luminosity. This analysis requires a systematic preliminary tune-up of the two rings involving optics, closed orbit, dispersion function and betatron coupling [17].

The betatron coupling as measured at the synchrotron light monitor in single beam operation mode is $\kappa \sim 0.14\%$. This value is significantly lower than the one achieved during the past run which was in the range $\kappa = 0.2 \div 0.3\%$ and proves that the IR coupling compensation mechanism works properly. However it does not grant that the coupling at the IP is compensated as required, and therefore it is measured by a beam-beam scan.

Luminosity has been studied by storing 100 bunches in collision at low current (less than 1 mA per bunch). Beam-beam overlap has been tuned independently in the transverse and longitudinal space using a fast monitor based on beam-beam bremsstrahlung [18]. The transverse overlap is performed by means of closed orbit bumps changing independently the position and angle of one beam through the opposite one. High resolution (0.5 μm) bumps are implemented by using four correctors for each ring; two of them are the DHCs already used to define the IR layout. Steering magnets are installed in the IR branches where the two rings are separated, but the coupling due to the detector solenoid is not compensated. As a consequence there is a cross-talk between closed orbit bumps in the two main betatron planes. In order to minimize perturbation on the beam trajectory and to avoid large kick values it has been decided to rotate the four correctors by an angle suitable to achieve horizontal closed orbit bumps without any additional vertical steering. The last issue is intended to preserve the vertical overlap, which is very demanding since DAΦNE works with flat beams. The convoluted vertical size of the colliding bunches at the IP measured by vertical beam-beam scan is $\sigma_y = 3.0$ μm, see Fig. 9, 14% lower than the best measured in the past after a long optimization period and with a beam optics having the same $\beta^*_y$ and natural emittance. The single bunch specific luminosity, defined as the single bunch luminosity divided by the product of the single bunch currents, at low currents, is of the order of $\sim 4.5 \cdot 10^{28}$ cm$^{-2}$ s$^{-1}$ the same as the one measured during the *Crab-Waist* test without detector solenoid.

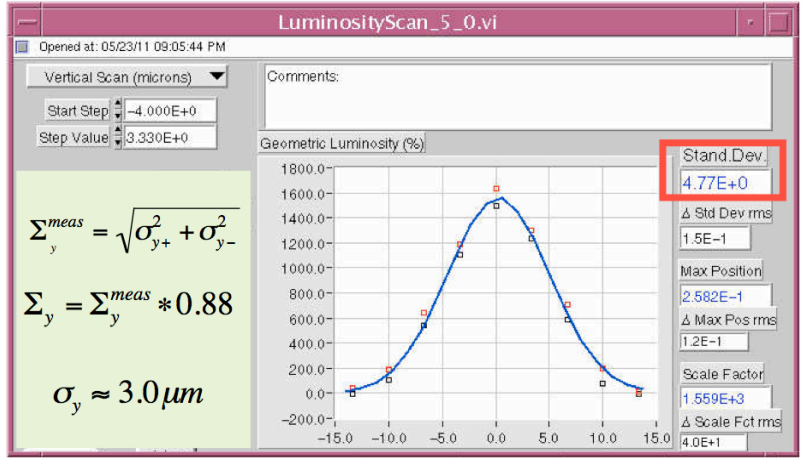

Figure 9. Vertical beam-beam luminosity scan.

## 4. Conclusions

A high luminosity IR compatible with a large detector having an intense solenoidal field has been designed, built and installed on the DAΦNE collider. All the different aspects related to layout, beam acceptance, optics and coupling correction have been studied in detail and optimized. Transverse coupling correction is achieved by an innovative scheme independently tunable for the two beams. The IR optics has been defined in order to fulfill all the requirements in terms of low beta functions at the IP, coupling correction and inclusion of the CW sextupoles. Tracking studies have been performed to ensure compatibility of the new IR structure with the mechanical layout of the ring arcs. The IR vacuum chamber mechanical design meets the requirements of lower impedance and larger free solid angle set by the collider and the experimental detector respectively. Test with the beams in collision have confirmed the effectiveness of the design criteria in minimizing transverse betatron coupling and providing high specific luminosity at low current.